%% file: arxiv211110.tex
\documentclass{llncs}
\usepackage{makeidx}         % allows index generation
\usepackage{graphicx}        % standard LaTeX graphics tool
\usepackage{multicol}        % used for the two-column index
\usepackage[bottom]{footmisc}% places footnotes at page bottom
\makeindex             % used for the subject index
\usepackage{epsfig, psfig,latexsym}
\usepackage{amsmath}
\usepackage{amsfonts}
\usepackage{graphics}
 \usepackage{amssymb}
 \usepackage{bbm}
 \usepackage{times}
\usepackage{amstext}
\usepackage{theorem}
\usepackage{proof}
\input{tempdef}

\begin{document}

\newcommand{\last}{\bigodot}
\newcommand{\alwaysP}{\blacksquare}
\newcommand{\sometimesP}{\bla\documentclass{llncs}cklozenge}
\newcommand{\LTLL}{$LTL_{\lambda=}\;$}
\newcommand{\Qu}{\mathbb Q}
\newcommand{\Zed}{\mathbb Z}
\newcommand{\Complex}{\mathbb C} 
\newcommand{\Real}{\mathbb R}

\title{Finite Model Finding for Parameterized Verification}
\author{Alexei Lisitsa}

%\\Department of Computer Science \\ the University of Liverpool \\
%Ashton Building, Ashton Street,\\ 
%Liverpool, L69 7ZF, U.K. 
%\texttt{alexei@csc.liv.ac.uk}}
\institute{Department of Computer Science\\
University of Liverpool, UK\\
%\\ 
%Ashton Building, Ashton Street,\\ 
%Liverpool, L69 7ZF, U.K.\\
\texttt{A.Lisitsa@liverpool.ac.uk}
}

\maketitle

\begin{abstract}
%{\bf rewrite}

In this paper we investigate to what extent  a very simple and natural "reachability as deducibility" approach, 
originating in  research on formal methods for security,  
is applicable to the automated verification of large  classes of infinite state and parameterized systems.   
This approach  is based on modeling the reachability between (parameterized) states as deducibility between 
suitable encodings of states by formulas of first-order predicate logic. The verification of a safety property 
is reduced  to the purely logical problem of finding a countermodel for a first-order formula. 
This task is then delegated then  to  generic automated finite model building procedures.  
In this paper we first establish the relative completeness of the finite countermodel finding method (FCM) for a class 
of parameterized linear arrays of finite automata. The method is shown to be at least as powerful as  known methods based on 
monotonic abstraction and symbolic backward reachability. 
%The verification of parameterized mutual exclusion 
%protocol using a finite model finder Mace4 is presented as a case study. 
Further, we extend the relative 
completeness of the approach and show that it can solve  all safety verification problems  which can 
be solved by regular model checking.  
%To illustrate the practical efficiency of 
%the approach we report on experiments  with  the  
%benchmarks of the infinite state and parameterized verification taken from the literature.     

\end{abstract}

%\vspace*{-10mm}

\section{Introduction}

The verification of infinite state systems  and parameterized systems is, in general, an undecidable algorithmic problem. 
That means the search for efficient procedures to tackle the larger and larger subclasses of verification tasks will never end. In this paper we 
investigate to what extent  a very simple and natural "reachability as deducibility" approach is applicable to the verification such systems. 
Consider an encoding $e: s \mapsto \varphi_{s}$ of states of a transition system ${\cal S} = \langle S,\rightarrow \rangle$   
by formulae of first-order predicate logic satisfying  the following property. The state  $s'$ is  reachable from $s$, i.e.  $s \rightarrow^{\ast} s'$ if and only if  $\varphi_{s'}$ is the logical 
consequence of $\varphi_{s}$, that is $\varphi_{s} \models \varphi_{s'}$ or $\varphi_{s} \vdash \varphi_{s'}$. 
%\footnote{Here we assume 
%standard definitions of semantic consequence $\models$ and deducibility $\vdash$ (in a complete deductive system)  for 
%the first-order predicate logic}. 
Under such assumptions 
%one can translate reachability questions for $S$ to the classical questions in logic. 
establishing reachability  amounts to theorem 
proving, while deciding non-reachability,  becomes theorem disproving. To verify a safety property, i.e 
non-reachability of \emph{unsafe} states, it is sufficient to  \emph{disprove} a formula of  the form 
$\phi \rightarrow \psi$. 
%To disprove $\phi \rightarrow \psi$ one can demonstrate a countermodel for it, i.e. a 
%model for $\phi \land \neg \psi$
Also,  in the case of safety verification already half of the assumption on the encoding is suffcient: $(s \rightarrow^{\ast} s') \Rightarrow 
(\varphi_{s} \vdash \varphi_{s'})$.  
The task of disproving can be delegated then to generic model finding  
procedures for first-order predicate logic \cite{Model}. 

Such an approach to verification originated within  research on formal methods for the analysis of 
cryptographic protocols \cite{Weid99,S01,GL08,JW09}.  
Being unaware of these developments in the verification of cryptographic protocols and coming from 
a different perspective  we re-invented the finite countermodel finding approach and 
applied it in a different context of verification of parameterized and infinite state systems 
\cite{Avocs09,AL09,AL10}.

%Due to the fact that countermodels may be inevitably infinite for some unsatisfiable first-order formulae, it is clear, 
%the approach can not be universally applied. However, quite unexpectedly, it has turned out to be practically efficient 
%for many known classes of parametrized and infinite-state verification tasks. 
We show in \cite{AL10} that the parallel composition of a complete finite model finder and a complete theorem prover provides 
a decision procedure for safety properties of lossy channel  systems \cite{Ab} under appropriate encoding. 
Using a finite model finder, Mace4, \cite{McCune} coupled with a theorem prover Prover9 \cite{McCune} we successfully 
applied the method to the verification of alternating bit protocol, specified within a lossy channel system; 
all parameterized cache coherence protocols from \cite{Del03}; series of coverability and reachability 
tasks conserning  Petri Nets;  parameterized Dining Philosophers Problem (DPP) and to parameterized linear 
systems (arrays) of finite automata.

Despite the wide range of parameterized verification tasks being tackled successfully by the method, the only result concerning  
completeness  presented so far  is that on the verification of lossy channel systems \cite{AL10}. 
The aim of this paper is to investigate further the completeness of the finite countermodel finding method 
for much larger  classes of parameterized verification tasks. Note that we focus here on  \emph{relative} completeness with 
respect to  well-known methods. To introduce the method we present as  case 
study in Section~\ref{sec:case} the 
details of automated verification of a parameterized mutual exclusion protocol, which is an instance of parameterized model  
defined in Section~\ref{sec:linear}.  
Further, we  present  an appropriate translation  of verification tasks for the parameterized systems of finite automata arranged in 
linear arrays into  formulae of first-order predicate logic (subsection~\ref{subsec:fo}). 
%and  
%establish  the relative completeness of the countermodel finding method 
%for this  class of verification problems. 
We show, in subsection~\ref{subsec:comp}, that the proposed finite countermodel finding method is at least as powerful as the methods based on 
monotone abstraction and symbolic backward reachability analysis \cite{Mon} for this class of verification 
problems. 
Further, in Section~\ref{sec:rmc} we extend the relative 
completeness of the approach and show that it can solve  all safety verification problems  which can 
be solved by a traditional \emph{regular model checking} \cite{RMC}.  In Section~\ref{sec:rel} 
we discuss related work and Section~\ref{sec:con} concludes the paper.

%Such parameterized systems have already been approached with automated verifications methods e.g. in  \cite{Aut, Mon}. 

%In this paper we establish the relative completeness of the method for a class of parameterized linear arrays of finite automata, 
%with respect to known methods based on monotone abstraction and symbolic backward reachability. 
%The verification of parameterized mutual exclusion protocol using a finite model finder Mace4 is presented as a case study.

\subsection{Preliminaries}~\label{sec:pre}
We assume that the reader is familiar with the the basics of first-order logic and algebra. In particular, we use without definitions the 
following concepts: first-order predicate logic, first-order models, interpretations of relational, functional and constant symbols, 
satisfaction $M \models \varphi$ of a formula $\varphi$  in a model $M$, semantical consequence $ \varphi\models\psi$, deducibility (derivability) $\vdash$ in first-order logic,    
monoid, homomorphism, finite automata and the  algebraic characterization of regular languages. We denote interpretations  by square brackets, so, 
for example, $[f]$ denotes an interpretation of a functional symbol $f$ in a model.
We also use the existence of \emph{complete} finite model finding procedures for the first-order predicate 
logic \cite{Model,McCune}, which given a first-order sentence $\varphi$  eventually produce a finite model for $\varphi$ if such a model exists.

%When the safety is verified, the method produces a finite countermodel, which is a concise representation of a 
%system  \emph{invariant}. We discuss the invariants produced for some of the mentioned examples, fthe ocussing on the one case study. This case study is the verification of parameterized mutual exclusion protocol, which was used as a running example in \cite{Mon}. The protocol is specified as a parameterized system of finite automata arranged in the linear array. 
%The transitions of automata, in general,   depend on \emph{global state} of the system. 
%Unlike previously considered cases, the first-order encoding of the protocol is designed in such a way that the 
%deducibility  models non only rechability by execution of the protocol, but also the \emph{construction}   of initial 
%states  
%for all partciular sizes of the system. In a sense, parameterized task is reduced to a signle infinite-state verification %task.

%We conclude with a general claim of relative completeness of the proposed method with  respect to the verification methods presented in \cite{Ab, Del03, Mon}. In the ongoing work we aim to formally support this claim.    

%Compact presentation of the paper contribution. 

\section{Parameterized linear arrays of automata}~\label{sec:linear}

The computational model we first consider in this paper consists of  parameterized systems of finite automata arranged in linear arrays 
\cite{Mon}. 
%Informally, the model works as follows. 
%Starting in some initial configuration the systems of automata evolve in the following way. 
%At every step one automaton is chosen non-deterministically to perfom a transition, which may be \emph{local}, in 
%which case it is performed unconditionally, or it may  have some \emph{global} conditions, in which case the global conditions are checked, and 
%if staisfied, the transition is performed.   
Formally, a parameterized system ${\cal P}$ is a pair $(Q,T)$, where $Q$ is a finite set of local states of processes and $T$ is finite set of transition rules. 
Every transition rule has one of the following forms

\begin{itemize}
\item $q \rightarrow q'$ where $q,q' \in Q$; 
\item ${\cal G}: q \rightarrow q'$, where $q,q' \in Q$ and ${\cal G}$ is a condition of the form $\forall_{I}J$, or $\exists_{I}J$ 
\end{itemize}

\noindent 
Here $J \subseteq Q$ and $I$ is an indicator of the context, and it may be one of the following: $L$ (for Left), $R$ (for Right), 
or $LR$ (for both Left and Right).

%No matter what context is, a formula $\Phi$ in the universal condition has the form  
%$\wedge_{i \in I} q_{i}$ for some $I \subseteq \{1, \ldots k\}$. Likewise, the formula $\Psi$ in the existential 
%condition has the form $\vee_{j \in J} q_{j}$ for some $J \subseteq \{1, \ldots k\}$.    

Given a parameterized system ${\cal P} = (Q,T)$ the \emph{configuration} of the system is a word 
$\bar{c} = c_{1}c_{2}\ldots c_{n} \in Q^{*}$. Intuitively, the configuration represents the local states of a family of 
$n$ finite state automata (processes) arranged in a linear array, so,  for example $c_{i} \in Q$ is a local state of automaton at position $i$ 
in the array.  

For a configuration $\bar{c} =c_{1}\ldots c_{n}$, position $i: 1 \le i \le n$ and a condition, we   define $\models$, a \emph{satisfaction} relation:  

\begin{itemize}
\item $(\bar{c},i) \models \forall_{L}J$ iff $\forall k < i \;\;  c_{k} \in J$;
\item $(\bar{c},i) \models \forall_{R}J$ iff $\forall k  > i \;\;  c_{k} \in J$; 
\item $(\bar{c},i) \models \forall_{LR}J$ iff $(\bar{c},i) \models \forall_{L}J$ and $(\bar{c},i) \models \forall_{R}J$
\item $(\bar{c},i) \models \exists_{L}J$ iff $\exists k < i \;\; c_{k} \in J$; 
\item $(\bar{c},i) \models \exists_{R}J$ iff $\exists k > i \;\; c_{k} \in J$;
\item $(\bar{c},i) \models \exists_{LR}J$ iff $(\bar{c},i) \models \exists_{L}J$ or $(\bar{c},i) \models \exists_{R}J$ 
\end{itemize} 

\noindent 
A parameterized system ${\cal P} = (Q,T)$ induces a transition relation $\rightarrow_{\cal P}$ on the set $C$ of all configurations as follows. For two configurations $\bar{c} \rightarrow_{\cal P} \bar{c'}$ holds iff either

\begin{itemize}
\item $q \rightarrow q'$ is a transtion in $T$ and for some $i: 1 \le i \le n$ $c_{i} = q$, $c_{i}'=q'$ and $\forall j \not=i \;\; c_{j} = c_{j}'$, or  
\item ${\cal G}: q \rightarrow q'$ is a transition in $T$ and for some $i: 1 \le i \le n$ $c_{i} = q$, $c_{i}' = q'$,  
$(\bar{c}, i) \models {\cal G}$ and  $\forall j \not=i \;\; c_{j} = c_{j}'$
\end{itemize}

\noindent 
The general form of the verification problem we consider here is as follows.  
\vspace*{3mm}

{\it \underline{Given:} A parameterized system ${\cal P} = (Q,T)$, a set $In \subseteq C$ of initial configurations, a set $B \subseteq C$ of \emph{bad} configurations.  

\underline{Question:} Are there any configurations $c \in In$ and $c' \in B$ such that 
$c'$ is reachable from $c$ in ${\cal P}$, i.e. for which  $c \rightarrow_{\cal P}^{\ast} c'$ holds? 
}

A negative answer for the above question means the safety property (``not B'') holds for the parameterized system. 
%We are interested here in the automated methods  for resolving such 
%questions. To explain the main ideas behind the proposed method we start with a case study.  
 
\vspace*{3mm}

\section{Case study}~\label{sec:case} 
\subsection{Mutual Exclusion Protocol}
We consider the verification of the parameterized mutual exclusion protocol which was used as an illustrative example in \cite{Mon}. This protocol is specified 
as a parameterized system ${\cal ME} = (Q,T)$, where $Q = \{green, black, blue, red\}$ and $T$ consists of the following transitions: 
\begin{itemize}
\item $\forall_{LR}\{green,black\}: green \rightarrow black$
\item $black \rightarrow blue$
\item $\exists_{L}\{black, blue, red \}: blue \rightarrow blue$
\item $\forall_{L}\{green\}: blue \rightarrow red$
\item $red \rightarrow black$
\item $black \rightarrow green$
\end{itemize}

The set of initial configurations $In = green^{\ast}$ consists of all configurations with all automata in $green$ states. 
The safety property we would like to check is a mutual exclusion of red states, i.e. in any reachable configuration, there are no more than one automaton 
in the $red$ state. 
The set $B$ of bad configurations is defined then by straightforward regular expression $B = Q^{\ast} \; red \; Q^{\ast} \; red \; Q^{\ast}$.     

\subsection{First-Order encoding}
We define a translation of the above parameterized system into a set of formulae $\Phi_{\cal P}$ of first-order logic. 
The vocabulary of $\Phi_{\cal P}$ consists of 
\begin{itemize}
\item constants $green$, $blue$, $black$, $red$ and $e$
\item one binary functional symbol $\ast$ 
\item unary predicates $R$, $G$, $GB$
\end{itemize}

\noindent
Given a configuration $\bar{c} = c_{1} \ldots c_{n}$ of ${\cal P}$ define its term translation as $t_{\bar{c}} = c_{1} \ast \ldots \ast c_{n}$.
It is well-defined modulo the associativity of $'\ast'$ which we will specify in the formula,  and uses an assumption that in the language we have  
all the elements of $Q$ as constants.   

The intended meaning of atomic formula $R(t_{\bar{c}})$ is that the configuration $\bar{c}$ is reachable, while $G(t_{\bar{c}})$ and $GB(t_{\bar{c}})$ mean 
$\bar{c}$ has only automata in $green$ states, and $\bar{c}$ has only automata in $green$ or $black$ states, respectively. 

Let $\Phi_{\cal P}$ be a set of the following formulae, which are all assumed to be universally closed: 

\begin{itemize}
\item $(x \ast y) \ast z  = x \ast (y \ast z)$
\item $ e \ast x = x \ast e = x$
\end{itemize} 
\begin{quote}
{\small ($\ast$ is a monoid operation and $e$ is a unit of a monoid)}
\end{quote}
\begin{itemize}
\item $G(e)$
\item $G(x) \rightarrow G(x * green)$
\end{itemize} 
\begin{quote}
{\small (specification of configurations with all $green$ states)}
\end{quote}
\begin{itemize}
\item $GB(e)$  
\item $GB(x) \rightarrow GB(x * green)$
\item $GB(x) \rightarrow  GB(x * black)$ 

\end{itemize} 
\begin{quote}
{\small (specification of configurations with all states being $green$ or $black$)}
\end{quote}
\begin{itemize}
\item $G(x) \rightarrow R(x)$
\end{itemize}
\begin{quote}
{\small (initial state assumption: ``allgreen'' configurations are reachable)} 
\end{quote}
\begin{itemize}
\item $(R((x * green) * y) \; \& \; GB(x)\; \& \; GB(y))\rightarrow R((x * black) * y)$
\item $R((x * black) * y) \rightarrow  R((x * blue) * y)$
 \item $R((x * blue) * y)\; \& \; (x = (z * black) * w) \rightarrow R((x * blue) * y)$ 
\item $R((x * blue) * y) \; \& \; (x = (z * blue) * w) \rightarrow R((x * blue) * y)$ 
\item $R((x * blue) * y) \;\&\; (x = (z * red) * w) \rightarrow R((x * blue) * y)$
\item  $R((x * blue) * y)\; \& \; G(x) \rightarrow R((x * red) * y)$
\item $R((x * red) * y) \rightarrow R((x * black) * y)$ 
\item $R((x * black) * y) \rightarrow R((x * green) * y)$  
\end{itemize} 

\begin{quote}
{\small(specification of reachability by one step transitions from $T$; one formula per transition, except the case with an existential condition, where three formulae
are used)}
\end{quote}
  
\noindent
Now we have a key proposition
 
\begin{proposition} [adequacy of encoding]
If a configuration $\bar{c}$ is reachable in ${\cal ME}$ then $\Phi_{\cal P} \vdash R(t_{\bar{c}})$
\end{proposition} 
{\bf Proof} By straightforward induction on the length of transition sequences in ${\cal ME}$ $\Box$ 

\subsection{Verification}

It follows now,  that to establish safety property of the protocol (mutual exclusion), it does suffice to show that
$\Phi_{\cal P} \not\vdash \exists x \exists y \exists z R((((x * red) * y) * red) * z)$. 
Indeed, if, on the contrary,  some bad configuration $\bar{c}$ would be reachable, then by Proposition 1 we would have 
for some terms $t_{1}, t_{2}, t_{3}$ that $\Phi_{\cal P} \vdash R(t_{\bar{c}})$ where 
$t_{\bar{c}} = (((t_{1} * red) * t_{2}) * red) * t_{3}$, and therefore $\Phi_{\cal P} \vdash\exists x \exists y \exists z R((((x * red) * y) * red) * z)$. Further, to show non-deducibility, it is sufficient to find a countermodel for 
$\Phi_{\cal P} \rightarrow  \exists x \exists y \exists z R((((x * red) * y) * red) * z)$.

Now we propose to delegate this last task to an automated procedure for finite model finding, which would search for a \emph{finite} model 
for $$\Phi_{\cal P} \land  \neg  \exists x \exists y \exists z R((((x * red) * y) * red) * z)$$ In the practical implementation of this scheme we 
used a finite model finder Mace4 \cite{McCune}, which was able to find a required model in 0.03 seconds. 
Actual input for Mace4 and further details  can be found in \cite{AL09}. 

A priori, to disprove some implication in first-order logic, searching for finite countermodels may be not sufficient, for such 
countermodels may inevitably be infinite. It has turned out empirically though that for many known parameterized (classes of ) problems, finite model 
finding is, indeed, both sufficient and efficient. In \cite{Avocs09} we established the first result on completeness of the method for 
a particular class of infinite state verification tasks. Here we demonstrate further results on \emph{relative} completeness.

\section{Correctness and Completeness}~\label{sec:corr}

\subsection{First-Order Encoding for General Case}~\label{subsec:fo} 
In the general form of the verification problem above we have to agree  what are the allowed sets of initial and bad configurations can be, and what are 
 their constructive representations.  Here we assume that 

\begin{itemize}
\item one of the local states $q_{0} \in Q$ is singled out as an initial state,  and the set $\mathit{Init}$ of initial configurations 
is always $q^{\ast}_{0}$, i.e. it consists of all configurations that have  all the automata in their local initial states;    
\item The set $B$ of bad configurations is defined  by a finite set  of words $F  \subseteq Q^{\ast}$: 
$B = \{ \bar{c} \mid  \exists \bar{w} \in F \; \land \; \bar{w} \preceq \bar{c}\}$, where $\bar{w} \preceq \bar{w}'$ denotes that 
$\bar{w}$ is a (not necessarily contiguous) subword of $\bar{w}'$. The elements of such $F$ are called \emph{generators} of  $B$. 
\end{itemize}   

\noindent
To illustrate this last point,  in our Case Study above, the set of bad configurations $B$ is defined by an $F$ consisting of one 
word with two symbols $red$ $red$. 

Given a parameterized system ${\cal P = (Q,T)}$, an intial local state $q_{0} \in Q$, a finite set of words $F$, we translate all of 
this into a set of formulae in first-order logic. 

The vocabulary consists of 
\begin{itemize}
\item constants for all elements of $Q$ plus one distinct constant, so we take  $Q \cup \{e \}$, with $e \not\in Q$ as the set of constants;  
\item the binary functional symbol $\ast$;
\item the unary relational symbol $\mathit{In}$;  
\item the unary relational symbol $R$; 
\item for every condition $\forall_{I} J$ in  the transitions from $T$ a unary relational symbol $P^{J}$ 
\end{itemize}

\noindent
Let $\Phi_{P}$ be the set of the following formulae, which are all assumed to be universally closed: 

\begin{itemize}
\item $(x \ast y) \ast z  = x \ast (y \ast z)$
\item $ e \ast x = x \ast e = x$
\end{itemize} 
%\begin{quote}
%{\small ($\ast$ is a monoid operation and $e$ is a unit of a monoid)}
%\end{quote}

\begin{itemize}
\item $In(e)$
\item $In(x) \rightarrow In(x * q_{0})$
\item $In(x) \rightarrow R(x)$
\end{itemize}

\noindent
For every condition $\forall_{I} J$ in the transitions from $T$:  
\begin{itemize}
\item $P^{J}(e)$
\item $wedge_{q \in J} (P^{J}(x) \rightarrow P^{J}(x * q))$  
\end{itemize}
  
\noindent
For every  unconditional transition $q_{1} \rightarrow q_{2}$ from $T$: 
\begin{itemize}
\item $R((x * q_{1}) * y) \rightarrow R((x * q_{2}) * y)$
\end{itemize}

\noindent
For every conditional transition $\forall_{L} J \;  (q_{1} \rightarrow q_{2})$ from $T$: 

\begin{itemize}
\item $(R((x * q_{1}) * y) \land P^{J}(x))  \rightarrow R((x *q_{2}) * y)$
\end{itemize}

\noindent
For every conditional transition $\forall_{R} J \;  (q_{1} \rightarrow q_{2})$ from $T$: 

\begin{itemize}
\item $(R((x * q_{1}) * y) \land P^{J}(y))  \rightarrow R((x *q_{2}) * y)$
\end{itemize}

\noindent
For every conditional transition $\forall_{LR} J \;  (q_{1} \rightarrow q_{2})$ from $T$: 

\begin{itemize}
\item $(R((x * q_{1}) * y) \land P^{J}(x) \land P^{J}(y))  \rightarrow R((x *q_{2}) * y)$
\end{itemize}

\noindent
For every conditional transition $\exists_{L} J \; (q_{1} \rightarrow q_{2})$ from $T$: 

\begin{itemize}
\item $\wedge_{q \in J} (R(x * q_{1}) * y) \land (x = (z * q) * w)) \rightarrow R((x * q_{2}) *y)$
\end{itemize}

\noindent
For every conditional transition $\exists_{R} J \; (q_{1} \rightarrow q_{2})$ from $T$: 

\begin{itemize}
\item $\wedge_{q \in J} (R(x * q_{1}) * y) \land (y = (z * q) * w)) \rightarrow R((x * q_{2}) *y)$
\end{itemize}

For every conditional transition $\exists_{LR} J \; (q_{1} \rightarrow q_{2})$ from $T$: 

\begin{itemize}
\item $\wedge_{q \in J} (R(x * q_{1}) * y) \land ((x = (z * q) * w) \lor (y = (z * q) * w))) \rightarrow R((x * q_{2}) *y)$
\end{itemize}

\noindent That concludes the definition of $\Phi_{\cal P}$. Next,  for a word $\bar{w} = w_{1}, \ldots, w_{n} \in Q^{\ast}$ we define (up to the 
associativity of $\ast$)  the formula  
$\psi_{\bar{w}}$ as R($x_{0} \ast w_{1} \ast x_{1} \ast \ldots \ast x_{n-1} \ast w_{n} \ast x_{n})$ where $x_{0}, \ldots x_{n}$ are variables. 
Finally, we define $\Psi_{F}$ as $\exists \bar{x} \vee_{\bar{w} \in F} \psi_{\bar{w}}$ (here we assume that all variables are bound by existential 
quantifiers).  

The following generalization of  Proposition 1 holds. 

\begin{proposition} [adequacy of encoding]
If configuration $\bar{c}$ is reachable in ${\cal P}$ then $\Phi_{\cal P} \vdash R(t_{\bar{c}})$
\end{proposition}
{\bf Proof} By straightforward induction on the length of the transition sequences $\Box$. 

\begin{corollary} [correctness of the method]
If $\Phi_{\cal P} \not\vdash \Psi_{F}$ then the answer to the question of the verification problem is negative, that is no bad configuration  is reachable from any of the initial configurations, and therefore, the safety property holds.
\end{corollary}

\subsection{Relative completeness}~\label{subsec:comp}
\noindent
Here we show that on the the class of the verification problems described above our proposed method is at least as powerful as the standard 
approach based on monotone abstraction \cite{Mon}.  Specifically,  if for  a parameterized system ${\cal P}$ the 
 approach \cite{Mon} proves a safety property, then our method based on finite countermodel finding will also succeed in establishing this property, provided a \emph{complete} 
finite model finding procedure is used.  

First, we briefly outline the monotone abstraction approach. Given a parameterized system ${\cal P} = (Q,T)$ and corresponding transition  
relation $\rightarrow_{\cal P}$  on the configurations withing ${\cal P}$, \cite{Mon} defines the monotonic abstraction $\rightarrow^{A}_{\cal P}$  of 
$\rightarrow_{\cal P}$ as follows. 
\begin{quote}
We have 
$\bar{c_{1}} \rightarrow^{A}_{\cal P} \bar{c_{2}}$ iff there exists a configuration $\bar{c}_{1}'$ such that $\bar{c}_{1}' \preceq \bar{c}_{1}$ and 
$\bar{c}' \rightarrow_{\cal P} c_{2}$. 
\end{quote}
Such defined $\rightarrow^{A}_{\cal P}$ is an over-approximation of $\rightarrow_{\cal P}$. 
To establish the safety property, i.e to get a  negative answer to the question of the verification problem above, \cite{Mon} proposes using   
a symbolic backward reachability algorithm  for monotone abstraction.  Starting with an upwards closed 
(wrt to $\preceq$) set of bad configurations $B = \{ \bar{c} \mid  \exists \bar{w} \in F \; \land \; \bar{w} \preceq \bar{c}\}$, the algorithm  
proceeds iteratively with the computation of the sets of configurations backwards reachable along $\rightarrow^{A}_{\cal P}$ from  $B$:  

\begin{itemize}
\item $U_{0} = B$ 
\item $U_{i+1} = U_{i} \cup Pre(U_{i})$
\end{itemize} 

where $Pre(U) = \{ \bar{c} \mid \exists \bar{c}' \in U \; \land \;   \bar{c} \rightarrow^{A}_{\cal P} \bar{c}'  \}$. 
Since the relation $\preceq$ is a \emph{well quasi-ordering} \cite{Mon} this iterative process is 
guaranteed to stabilize, i.e $U_{n+1} = U_{n}$ for some finite $n$. During the computation each   $U_{i}$ is  represented symbolically by a 
finite sets of generators. Once the process stabilized on some $U$ the check is performed on whether $\mathit{Init} \cap U = \emptyset$. If this 
condition is satisfied then the safety is established, for no bad configuration  can be reached from intial configurations 
via $\rightarrow^{A}_{\cal P}$ and, a fortiori, via $\rightarrow_{\cal P}$.        
 
\begin{theorem}[relative completeness]~\label{th:rc1}
Given a parameterized system ${\cal P} = (Q,T)$ and the set  of bad configurations 
$B = \{ \bar{c} \mid  \exists \bar{w} \in F \; \land \; \bar{w} \preceq \bar{c}\}$. Assume the algorithm described above terminates with 
$\mathit{Init} \cap U = \emptyset$. Then there exists  a \emph{finite model} for $\Phi_{\cal P} \land \neg \Psi_{F}$  
\end{theorem}
{\bf Proof.} First we observe that since $U \subseteq Q^{\ast}$ has a finite set of generators, it is a \emph{regular} set. According to the algebraic 
characterization  of regular sets, there exists a \emph{finite} monoid ${\cal M} = (M,\circ)$, a subset $S \subseteq M$ and a homomorphism 
$h : Q^{\ast} \rightarrow  {\cal M}$ from the free monoid $Q^{\ast}$ to ${\cal M}$ such that 
$U = \{ \bar{w} \mid \bar{w} \in Q^{\ast} \land h(\bar{w}) \in S \}$. We set $M$ to be domain of the required finite model. 

\noindent 
Now we define interpretations of constants: for $q \in Q$ $[q] = h(q)$ and $[e] = \underline{1}$, where $\underline{1}$ is an unit element of 
the monoid. 

\noindent
The interpretation $[\ast]$ of $\ast$ is a monoid operation $\circ$. We define an interpretation of $R$ as $[R] = M - S$.

\noindent
We define an interpretation of $In$ inductively: $[In]$ is the least subset of $M$ satisfying  
$\underline{1} \in [In]$ and $\forall x \in [In] \; x \circ [q_{0}] \in [In]$.  

An interpretation of $P^{J}$ is defined inductively as follows. $[P^{J}]$ is a least subset of $M$ satisfying 
$\underline{1} \in [P^{j}]$ and $\forall x \in [P^{J}]\; \forall q \in J \;  x \circ [q] \in  [P^{J}]$. 
That concludes the definition of the finite model, which we denote by $\gM$.  
The key property of the model is given by the following lemma. 
\begin{lemma}
$h(\bar{w}) \in [R]$ iff no bad configuration is $\rightarrow^{A}_{\cal P}$-reachable from $\bar{w}$. 
\end{lemma}

\noindent
{\bf Proof} is straightforward from  the definitions of  $U$, ${\cal M}$, $h$ and $[R]$. 

\vspace*{3mm}
\noindent 
It follows immediately  that  $\gM \models \neg \Psi_{F}$.  To show that $\gM \models \Phi_{\cal P}$ we  show that $\gM \models \varphi$ for every 
$\varphi \in \Phi_{\cal }$.  For the first seven formulae in the definition of $\Phi_{\cal P}$ this involves a routine check of definitions. 
We show here only one case of the remaining formulae axiomatizing $R$.

To demonstrate  
$\gM \models  (R((x * q_{1}) * y) \land P^{J}(x))  \rightarrow R((x *q_{2}) * y)$ for some $\forall_{L} J \;  (q_{1} \rightarrow q_{2})$ in $T$ assume that 
left-hand side of the implication is satisfied in $\gM$ for some assignment of the variables.  That means there are $t_{1}$, $t_{2} \in M$ such that $t_{1} \ast h(q_{1}) \ast t_{2} \in [R]$ and 
$t_{1} \in [P^{J}]$.  Furthermore, there are $\bar{w_{1}}, \bar{w_{2}} \in Q^{*}$ such that $t_{1} = h(\bar{w_{1}})$,  $t_{2} = h(\bar{w_{2}})$ and 
no bad states  are $\rightarrow^{A}_{\cal P}$-reachable  from $w_{1}\;q_{1}\;w_{2}$. Now,   transition by the rule 
$\forall_{L} J \;  (q_{1} \rightarrow q_{2})$  is possible from  $\bar{w_{1}}\;q_{1}\;\bar{w_{2}}$, resulting in  the configuration 
$\bar{w_{1}}\;q_{2}\;\bar{w_{2}}$, from which it is still the case that no bad configurations are reachable. This implies $h(\bar{w_{1}})\;h(q_{2})\;h(\bar{w_{2}}) \in [R]$, 
and therefore $\gM \models  (R((x * q_{1}) * y) \land P^{J}(x))  \rightarrow R((x *q_{2}) * y)$. 
The remaining cases are tackled  in a similar way. $\Box$.     
%It is straightforward now to check that such a model satisfies 
%$\Phi_{\cal P} \land 
%$\neg \Psi_{F}$. Indeed, assume 

%By definition of $[R]$ if $h(\bar{w}) \in [R]$ then no bad configuration is $\rightarrow^{A}_{\cal P}$-reachable from $\bar{w}$.   

 \subsection{FCM is stronger than monotone abstraction}

For some parameterized  systems the method based on monotone abstraction may fail to establish safety even though it may 
actually hold. The reason for this is a possible \emph{overapproximation} of the set of reachable states as a result of 
abstraction.  A simple example of such a case is given in \cite{context09}. The parameterized system $(Q,T)$ where
$Q = \{q_{0}, q_{1}, q_{2}, q_{3},q_{4}\}$ and where $T$ includes the following transition rules 
\begin{enumerate}
\item $\forall \{q_{0},q_{1},q_{4}\}: q_{0} \rightarrow q_{1}$
\item $q_{1} \rightarrow q_{2}$
\item $\forall_{L}\{q_{0}\}: q_{2} \rightarrow q_{3}$
\item $q_{3} \rightarrow q_{0}$
\item $\exists_{LR}\{q_{2}\}: q_{3} \rightarrow q_{4}$
\item $q_{4} \rightarrow q_{3}$ 
\end{enumerate}

\noindent 
satisfies  mutual exclusion for  state $q_{4}$, but this fact can not be established by the monotone abstraction method from 
\cite{Mon}. However, using first-order encoding presented above and the finite model finder we have verified mutual exclusion for 
this system,  demonstrating that FCM method is stronger than monotone abstraction.  
Mace4 has found  a finite countermodel of the size 6 in 341s. See details in \cite{AL09}  and the Appendix. 

The issue of overapproximation has been addressed in \cite{context09} where  two  refinements of the monotonic abstraction  
method were proposed. One   resulted in an exact context-sensitive symbolic algorithm which allows one to compute exact 
symbolic representations of predecessor configurations, but the termination of which is not guaranteed. On the other hand, an   
approximated context-sensitive symbolic algorithm is  also proposed and while  guaranteed to terminate, may still lead to overapproximation.       
One can show the relative completeness of the FCM method with respect to both algorithms for the case of safety verification. 
In both algorithms the safety is established when a finite representation of a set $U$  of configurations backwards rechable from 
unsafe states,  is obtained upon an algorithm termination. In both cases such a set $U$ can be shown is regular, and therefore 
one can apply the arguments used in the proof of Theorem~\ref{th:rc1}. We postpone the detailed presentation 
till another occassion, but would like to emphasize that  the main reason for the relative completeness here is a mere \emph{existence} 
of the  regular  sets of configurations  subsuming all reachable configurations and disjoint with unsafe configurations.

\section{Regular model checking}~\label{sec:rmc}  

The result of the previous section may appear  rather narrow  and related to  a specific  class 
of parameterized systems. The verification of safety for this class can be re-formulated for,   and dealt
with the traditional  \emph{regular model checking} approach\cite{RMC}. 
In this section we extend our relative completeness result 
% considerably
and show that whenever safety for a 
parameterized system  can be established by the regular model checking approach then it can also be verified 
by the finite countermodel finding method.

We start with the basics of the traditional regular model checking approach, borrowing  standard definitions  
largely from  \cite{HV07}. 
A finite automaton is a tuple $M = \langle Q, \Sigma, \delta, q_{0}, F \rangle$, where   $Q$ is a finite set of states, $\Sigma$ is a finite 
alphabet, $\delta \subseteq Q \times \Sigma \times Q$ is a set of transitions, $q_{0} \in Q$ is an initial state 
and $F \subseteq Q$ is a set of final (accepting) states. 
$M$ is deterministic automaton if $\forall q \in Q\; \forall a \in \Sigma$ there esists at most  one $q'$ such that $\langle q,a,q'\rangle \in \delta$. 
With every finite automaton we associate a transition relation $\rightarrow \; \subseteq Q \times \Sigma^{\ast} \times Q$ which is defined 
as the smallest relation satisfying: (1): $ \forall q \in Q; q \rightarrow^{\epsilon}q$, (2) if $\langle q,a,q'\rangle \in \delta$, then 
$q \rightarrow^{a} q'$, (3) if $q \rightarrow^{w} q'$ and $q' \rightarrow^{a} q''$ then $q \rightarrow^{wa} q''$. 
The language recognized by the automaton $M$ is defined as 
$L(M) = \{w \mid \exists q' \in F \; \land \; \land \;  q_{0} \rightarrow^{w} q'\}$. 

Let $\Sigma$ be a finite alphabet and $\epsilon \not\in \Sigma$. Let $\Sigma_{\epsilon}$ = $\Sigma \cup \{\epsilon\}$. 
A \emph{finite transducer} over $\Sigma$ is a tuple $\tau = \langle Q, \Sigma^{\ast}_{\epsilon} \times \Sigma^{\ast}_{\epsilon}, \delta, q_{0}, F \rangle$, where $Q$ is a finite set of states, $\delta \subseteq Q \times \Sigma_{\epsilon} \times \Sigma_{\epsilon} \times Q$ a set of transitions, $q_{0} \in Q$ is an initial state, and $F \subseteq Q$ is a set of final (accepting) states.  The transition relation $\rightarrow \subseteq Q \times \Sigma^{\ast} \times \Sigma^{\ast} \times Q$ is defined as the smallest relation staisfying: (1) $q \rightarrow^{\epsilon, \epsilon} q$ for every $q \in Q$, 
(2) if $\langle q,a,b,q' \rangle \in \delta$, then $q \rightarrow^{a,b}q'$, and (3) if $q \rightarrow^{w,u} q'$ and $q' \rightarrow^{a,b} q''$, 
then $q \rightarrow^{wa,ub} q''$. With every transducer $\tau$ we associate a binary  
relation $r_{\tau}= \{\langle w,u \rangle \mid \exists q' \in F \;\land\; q_{0} \rightarrow^{w,u} q' \}$. Let $r_{\tau}^{\ast}$ 
denote the reflexive and transitive closure of $r_{\tau}$.

The verification of safety properties in the framework of regular model checking proceeds as follows. The set of initial states  of the 
parameterized (or infinite state) system is presented by an effectively given (by a finite automaton)  
\emph{regular} language $\mathit{Init}$. The set  of ``bad", or unsafe states is described  by another regular language $Bad$. 
One-step transitions of the system to be verified are presented by a transducer relation $r_{\tau}$ (for some finite state transducer $\tau$). The verification of safety property (``never get into the bad states'') is reduced to the following 

\begin{problem}\label{problem:rmc}
Given regular sets $\mathit{Init}$ and $Bad$ and a finite transducer $\tau$, does $r_{\tau}^{\ast}(Init) \cap Bad = \emptyset$ hold? 
\end{problem}  

\noindent
Regular model checking (RMC)  is one of the most general methods for formal verification of parameterized and infinite 
state systems \cite{RMC,RMCs}. One of the issues with the method is that the termination of  the computation of 
transitive closure $r_{\tau}^{\ast}(Init)$ is not guaranteed.  
To alleviate this issue, various acceleration methods have been proposed. 
We show that the finite countermodel finding method is actually as powerful as any variant of RMC, 
the only assumption  to guarantee its termination is the \emph{existence} of a regular set $R$ subsuming  
$r_{\tau}^{\ast}(Init)$ and being disjoint with $Bad$.    

\subsection{From regular model checking to first-order disproving}
In this subsection we show how to reduce the generic regular model checking question posed in   
the Problem~\ref{problem:rmc} above to the problem of disproving of a formula from  classical first-order predicate logic. 
Solution of the latter problem is then delegated  to the generic automated finite model finding procedure.  

\medskip
\noindent
Assume we are given   

\begin{itemize}
\item a finite state automaton $M_{1} = \langle Q_{1}, \Sigma, \delta_{1},q_{{0}_{1}}, F_{1}\rangle$ recognizing a regular language $Init$;   
\item a finite state automaton $M_{2} = \langle Q_{2}, \Sigma, \delta_{2},q_{{0}_{2}}, F_{2}\rangle$ recognizing a regular language $Bad$; 
\item a finite state length-preserving transducer $\tau = \langle Q, \Sigma^{\ast} \times \Sigma^{\ast}, \delta, q_{0}, F \rangle$ representing the transition relation $r_{\tau}$; 
\end{itemize}

\noindent
Assume also (without loss of generality) that sets $Q_{1},Q_{2},Q,\Sigma$ are disjoint.

\noindent Now define a set of formulae of first-order predicate logic as follows. In fact, it is a formalization of the above 
definition of $\rightarrow$ within  first-order predicate logic. 

The vocabulary consists of 
\begin{itemize}
\item constants for all elements of  $\Sigma \cup Q_{1} \cup Q_{2} \cup Q$ plus one distinct constant $e$;  
\item a binary functional symbol $\ast$;
\item unary relational symbols $R$, $Init$ and $Bad$;   
\item a binary relational symbol $Trans$; 
\item a ternary relational symbol $T^{(3)}$; 
\item a 4-ary relational symbol $T^{(4)}$;
%\item for every condition $\forall_{I} J$ in  the transitions from $T$ a unary relational symbol $P^{J}$ 
\end{itemize}

\noindent
Let $\Phi$ be the set of the following formulae, which are all assumed to be universally closed:

\begin{enumerate}

\item $(x \ast y) \ast z = x \ast (y \ast z)$

\item $T^{(3)}(q,e,q)$ for all $q \in Q_{1} \cup Q_{2}$; 

%\item $T^{(3)}(x,e,x)$ (a variant of above);  

\item $T^{(3)}(q,a,q')$ for all $(q,a,q') \in \delta_{1} \cup \delta_{2}$; 

\item $T^{(3)}(x,y,z) \land T^{(3)}(z,v,w) \rightarrow T^{(3)}(x,y \ast v,w)$
 
\item $\vee_{q \in F_{1}} T^{(3)}(q_{0_{1}},x,q) \rightarrow Init(x)  $

\item $\vee_{q \in F_{2}} T^{(3)}(q_{0_{2}},x,q) \rightarrow Bad(x) $

\item $T^{(4)}(x,e,e,x)$

\item $T^{(4)}(q,a,b,q')$ for all $(q,a,b,q') \in \delta$

\item $T^{(4)}(x,y,z,v) \land T^{(4)}(v,y',z',w) \rightarrow T^{(4)}(x,y * y', z * z',w)$

\item $Trans(x,y) \leftrightarrow \vee_{q \in F} T^{(4)}(q_{0},x,y,q)$

\item $Init(x) \rightarrow R(x)$

%$R(e)$. 

\item $R(x) \land Trans(x,y) \rightarrow R(y)$

\end{enumerate}

\begin{proposition}[adequacy of Init and Bad translations]~\label{prop:InitBad}
 
\begin{itemize}
\item[] If $w \in Init$ then  $\Phi \vdash Init(t_{w})$
\item[] If $w \in Bad$ then $\Phi \vdash Bad(t_{w})$
\end{itemize}
\end{proposition}

\noindent {\bf Proof} For $w = s_{1}, \ldots s_{n}  \in Init$ we have $w$ is 
accepted by the finite automaton $M_{1}$, which means there is a sequence of states 
$q_{{0}_{1}}, q_{1}, \ldots q_{n}$ with $q_{n} \in F_{1}$ such that 
$\langle q_{i},s_{i},q_{i+1} \rangle \in \delta_{1}$ for $i = 0, \ldots n-1$. 
By the definition of $\Phi$ (clause 3) all formulae $T(q_{i},s_{i},q_{i+1})$ are in $\Phi$. 
Together with clause 4, this  gives   $\Phi \vdash T(q_{{0}_{1}},t_{w},q_{n})$. This  with $q_{n} \in F_{1}$ and 
using clause 5 entails $\Phi \vdash Init(t_{w})$. 
The second statement is proved in the same way. $\Box$

\begin{proposition}[adequacy of encoding]~\label{prop:adequacy}
If $w \in r^{\ast}_{\tau}(Init)$ then   $\Phi \vdash R(t_{w})$
\end{proposition}

\noindent {\bf Proof.}$\;$ Easy induction on the length of transition sequences. 

\begin{itemize}
\item\emph{Induction Base Case.} Let $w \in Init$. 
Then $\Phi \vdash Init(t_{w})$ (by Proposition~\ref{prop:InitBad}), and, further, $\Phi \vdash R(t_{w})$ 
(using clause 11).  

\item\emph{Induction Step Case.} Let $w \in r^{n+1}_{\tau}(Init)$. Then there exists $w'$ such that 
$w'\in r^{n}_{\tau}(Init)$   and $\langle w',w\rangle \in r_{\tau}$ . 
By the induction assumption $\Phi \vdash R(t_{w'})$. 
Further, by an argument analogous to the proof in Proposition~\ref{prop:InitBad}, 
$\langle w',w\rangle \in r_{\tau}$ entails $\Phi \vdash T(q_{0},t_{w'},t_{w},q)$ for some $q \in F$. 
It follows, using clause 10, that $\Phi \vdash Trans(t_{w'},t_{w})$.  From this, the clause 12 and the induction assumption 
$\Phi \vdash R(t_{w'})$ follows. 

\end{itemize}

\begin{corollary}\label{prop:corr2}
If $r^{\ast}_{\tau}(Init) \cap Bad \not=\emptyset$ then $\Phi \vdash \exists x (R(x) \land Bad(x))$.  
\end{corollary} 

\noindent
The Corollary~\ref{prop:corr2} serves as a formal underpinning of the proposed verification method. In order 
to prove safety, that is $r^{\ast}_{\tau}(Init) \cap Bad =\emptyset$ it  suffices to demonstrate 
$\Phi \not\vdash \exists x (R(x) \land Bad(x))$, or equivalently, to \emph{disprove} 
$\Phi \rightarrow \exists x (R(x) \land Bad(x))$. We delegate this task to the finite model 
finding procedures, which search for the \emph{finite} countermodels for 
$\Phi \rightarrow \exists x (R(x) \land Bad(x))$.

%\begin{example}  
%Verification of token passing protocol
%\end{example}

\subsection{Relative completeness with respect to  RMC}

As highlighted earlier, searching for \emph{finite} countermodels to disprove non-valid  
first-order formulae may not always  lead to success, because  for some formulae countermodels are 
inevitably \emph{infinite}. In this subsection we show that it is not 
the case for the first-order encodings of the 
problems which can be positively answered by RMC, and therefore such problems  can also be resolved positively 
by the  proposed finite countermodel finding method, 
provided a complete finite model finding procedure is used.   

% and in such  

%It has turned out empirically that for many parameterized, or infinite-state 
%verification problems (under appropriate encoding) searching for finite models is sufficient and, 
%in many cases,  practically efficient \cite{AL09, Avocs09, AL10}.  

%In \cite{AVOCS09,ATVA10} we proved that parallel composition of complete finite model finding procedure and 
%complete theorem prover for first-order predicate logic provides with the 
%decision procedure for \emph{lossy channel systems}. 
%In \cite{FMCAD10} we show the \emph{relative completeness} of the countermodel finding  based verification procedure for a class of parameterized linear automata arrays (class taken from \cite{Mon})  with respect to the verification procedure 
%based on monotone abstraction and symbolic reachability \cite{Mon}. 

%In this subsection we extend (relative) completeness results to the case of problems which can be 
%tackled by regular model checking. 

%More precisely, we show that if the verification problem, an instance of \emph{Problem 1}, 
%\begin{problem}\label{problem:rmc}{problem:rmc}
%Given regualr sets $Init$ and $Bad$ and a finite transducer $\tau$, does $r_{\tau}^{\ast}(Init) \cap Bad = \emptyset$ hol
%\end{problem}  
%can be resolved positively by the traditional regular model checking, then it can be resolved positively by the 
%proposed finite countermodel finding method too, provided the \emph{complete} finite model finding procedure is used. .  

%establishing thereby the relative completeness of the method. 

Assume that RMC answers positively the question of  
Problem~\ref{problem:rmc} above. 
In the RMC approach the positive answer follows from producing a \emph{regular} set ${\cal R}$ 
such that $r_{\tau}^{\ast}(Init) \subseteq {\cal R}$ and ${\cal R} \cap Bad = \emptyset$. 
We show that in such a case there always exists a \emph{finite} countermodel for $\Phi \rightarrow \exists x (R(x) \land Bad(x))$.

\begin{theorem}[relative completeness]~\label{th:rc2}
Let $Init$ and $Bad$ be  regular sets given by recognizing finite automata $M_{1}$ and $M_{2}$,   and $\tau$ be a finite state transducer. Let $\Phi$ be a first-order formula defined above. If there exists a \emph{regular} set ${\cal R}$
such that  $r_{\tau}^{\ast}(Init) \subseteq {\cal R}$ and ${\cal R} \cap Bad = \emptyset$ then there exists  a finite countermodel  
for $\Phi \rightarrow \exists x (R(x) \land Bad(x))$  
\end{theorem}    

\noindent 
{\bf Proof}

\noindent Since ${\cal R}$ is regular, according to the algebraic characterization of regular sets, 
there exists a finite monoid ${\cal M} = (M, \circ)$, a subset $S \subseteq M$ and a 
homomorphism $h: \Sigma^{\ast} \rightarrow {\cal M}$ such that 
${\cal R} = \{\bar{w} \mid \bar{w} \in \Sigma^{\ast} \land h(\bar{w}) \in S\}$. 

We take $M \cup Q_{1} \cup Q_{2}$ to be domain of the required  finite model, and then  
define interpretations as follows.   

\begin{itemize}
\item[$\bullet$] For $a \in \Sigma$ $[a] = h(a)$;   
\item[$\bullet$] $[e] = \underline{1}$, where $\underline{1}$ is an unit element of the monoid;  
\item[$\bullet$] $[\ast]$ is a monoid operation $\circ$; 
\item[$\bullet$] Interpretations of $T^{3}$ and $T^{4}$ are defined inductively, as the least subsets of tuples 
satisfying, respectively,  formulae (2)-(4) and (7) - (9)(and assuming all interpretations given above); 
\item[$\bullet$] Interpretations of $\mathit{Init}$ and $\mathit{Bad}$ are defined to be the least 
subsets satisfying (5) and (6), respectively (assuming all interpretations above); 
\item[$\bullet$] Interpretation of Trans is defined by (10)(assuming all interpretations above); 
\item[$\bullet$] Interpretation of $R$ is $S$.  
\end{itemize} 

Now it is straightforward to check that such defined a \emph{finite} model indeed
satisfies $\Phi \land \neg \exists x (R(x) \land Bad(x))$. Checking that $\Phi$ is satisfied is by routine 
inspection of  the definitions. To check that $\neg \exists x (R(x) \land Bad(x))$ is satisfied, 
assume the opposite holds.  So  there exists an element $a$  of the monoid ${\cal M}$ such that $a \in [R]$ and  
$a \in [Bad]$. Then, for a word $w \in \Sigma^{\ast}$ such that $h(w) = a$, we have $w \in {\cal R} \cap Bad \not=\emptyset$, which  
contradicts with the assumption of the theorem. $\Box$.

\subsection{Optimizations}~\label{sec:optimization} 

%The definition of $\Phi$ is not optimal. 
%For example, the predicate $Init$ was introduced just for transparency of presentation of main ideas. 
%One can get rid of this predicate just by deleting the  clause 11 in the list of formulae and replacing the occurrence of 
%$Init(x)$ in the clause 5 by the occurrence of $R(x)$. 
%It is easy to see that Proposition~\ref{prop:adequacy}, Corollary~\ref{prop:corr2} and Theorem~\ref{th:rc2} remain true after such a modification. 

%More substantially, 
\noindent
In many cases (i.e. in many subclasses of verification tasks),  
the transition relation and/or  the sets of `initial' and `bad' states are described  
not by finite state transducers/automata, but in more explicit and simpler ways, e.g. by 
 rewriting rules for transitions and simple grammars  generating sets of states. 
In such a cases,  first-order translations can be made simpler and the whole procedure more efficient. 
Our treatment of the case of parameterized linear automata in Section~\ref{sec:corr} can be seen as an illustration of 
such a modification.

\subsection{Experimental results}~\label{sec:exp}

\noindent 
In the experiments we used the finite model finder Mace4\cite{McCune} within the package 
Prover9-Mace4, Version 0.5, December 2007. 
It is not the latest available version, but it provides with convenient GUI for both the theorem prover and the finite 
model finder. The system configuration used in the experiments:   Microsoft Windows XP Professional, Version 2002, Intel(R) Core(TM)2 Duo CPU, 
T7100 @ 1.8Ghz  1.79Ghz,  1.00 GB of  RAM.  The time measurements are done by Mace4 itself, upon completion 
of the model search it communicates the CPU time used. The table below lists the parameterised/infinite state 
protocols together with the references and shows the time it took Mace4 to 
find a countermodel and verify a safety property.  
The time shown is an average of 10 attempts.

\begin{center}
  \begin{tabular}{| l | c | r | }
    \hline
    Protocol & Reference & Time \\ \hline
    Token passing (non-optimized)  & \cite{HV07} & 0.12s \\ \hline
    Token passing  (optimized) & \cite{HV07} & 0.01s \\
    \hline
    Mutual exclusion I & \cite{Mon} and 3 & 0.03s \\
\hline
Mutual exclusion II & \cite{context09} and 4.2 & 341s\\
\hline
    Bakery & \cite{RMC} & 0.03s \\
\hline 
Paterson$^{-}$ & \cite{beyond} and \ref{sec:beyond} & 0.77s\\
\hline
  \end{tabular}
\end{center}

\subsection{Beyond regular model checking}~\label{sec:beyond} 
The method of verification via disproving (countermodel finding) can be applied also to classes of problems where 
traditional regular model checking is not applicable. Consider, for example,  the case where the set of initial 
states is not regular, so the standard algorithms of RMC are not applicable. 
In the paper \cite{beyond} an extension of regular model checking is proposed, which is capable to tackle some non-regular cases. 
Not claiming any kind of completeness (yet!) we show in this subsection that a case study example from \cite{beyond} can be 
(partially, as for now) tackled by the finite countermodel finding method too. 
Consider the following string rewriting system over alphabet $\{0,1\}$, which is an encoding of the parameterized Paterson mutual exclusion algorithm from 
\cite{beyond}:
\begin{enumerate}
\item $x01y \rightarrow x10y$   $\;\;\;\;\;\;\;\;$   where $x \in 0^{\ast}$, $y \in (1 + 0)^{\ast}$
\item $x101y \rightarrow x110y$  $\;\;\;\;\;$  where $x \in (1 + 0)^{\ast}$, $y \in 1^{\ast}$
\item $x001y \rightarrow x010y$  $\;\;\;\;\;$  where $x, y \in (1 + 0)^{\ast}$
\item $x0 \rightarrow 0x$    $\;\;\;\;\;\;\;\;\;\;\;\;\;\;\;\;$      where $x \in (1+0)^{\ast}$
\end{enumerate}

The safety condition for this rewriting system is  `Starting from any string of the form 
$0^{n}1^{n}$ no string from the set $(0+1)^{\ast}00$  is reachable' (mutual exclusion of the original Paterson algorithm).  
In \cite{beyond} it is shown that the extension of RMC proposed there can successfully verify the condition. 

Following the translation from the Section~\ref{sec:corr} we encode the string rewriting system 
into a first-order formula $\Phi$. 
Since the set of initial states is not regular, the formula contains a part specifying the generation   
of initial states by a context-free grammar: $R(e) \land R(x) \rightarrow R((0 \ast x) \ast 1)$.

In the experiments we failed to verify the correctness condition for the Paterson algorithm, however, 
for the reduced  string rewriting system Paterson$^{-}$, containing only the rules 1,2,4  we have verified safety condition above. 
Mace4 has found a finite countermodel of size 8 in $0.77s$. 
The details can be found in \cite{AL09}.

%\subsection{Beyond regular model checking}~\label{sec:beyond} 
%\noindent 
%The method of verification via disproving (countermodel finding) can be applied also to classes of problems where 
%traditional regular model checking is not applicable. Some partial experimental results on the application 
%FCM to non-regular cases are presented in Section \ref{sec:beyond_app} of the Appendix.  

\section{Related work}~\label{sec:rel}

As mentioned Section 1 the approach to verification using the modeling of protocol executions by 
first-order derivations and together with  countermodel finding for disproving was introduced within the research on the 
formal analysis of cryptographic protocols.  It can be traced back to the early papers by Weidenbach \cite{Weid99} and by 
Selinger \cite{S01}. In \cite{Weid99} a decidable fragment  of Horn clause logic has been identified for which resolution-based 
decision procedure has been proposed (disproving by the procedure amounts to the termination of saturation without producing a 
proof). It was also shown  that the fragment is expressive enough to encode cryptographic protocols 
and the approach has been illustrated by the automated verification of some protocols using the SPASS theorem prover. In \cite{S01}, 
apparently for the first time,  explicit building of finite countermodels has been proposed as a tool to establish correctness 
of cryptographic protocols. It has been illustrated by an example, where a countermodel was produced manually, and the
automation of the process has not been disscussed.  The later work by Goubault-Larrecq \cite{GL08} has shown how a countermodel produced 
during the verification of cryptographic protocols can be converted into a formal induction  proof. 
Also, in \cite{GL08} different approaches to  model building have been discussed and it was argued that  
an implicit model building procedure using alternating tree automata  is more efficient in the situations when no small 
countermodels exist. Very recently, in the paper \cite{JW09} by J.~Jurgens and T.~Weber, an extension of Horn clause logic was proposed and 
the soundness of a countermodel finding procedure for this fragement has been shown, again in the context of cryptographic 
protocol verification. Furthermore, in \cite{JW09} an approach to the verification of parameterized cryptoprotocols is proposed.

The work we reported in this paper differs from all the approaches mentioned previously in two important  aspects. 
Firstly, to the best of our knowledge,  none of the previous work addressed verification via countermodel finding applied 
outside of the  area of cryptographic protocols (that includes the most recent work \cite{Gut} we are aware of).
Secondly, the (relative)  completeness for the classes of verification tasks has not been addressed in  previous work. 

%Notice that in \cite{Weid99,GL08} the complete decision procedures for the fragments of first-order logic, 
%involving, as a part,  the consistency checking (countemodel building), were proposed. However, we don't see a feasible way to 
%derive  our completeness results by the reduction to \cite{Weid99,GL08}, for first-order translations we use in this paper do not 
%fit into the fragments from \cite{Weid99,GL08}.  

The encoding  of infinite state systems in first-order predicate logic is used  
in the MCMT  deductive symbolic model checker \cite{MCMT,GNRZ}.
While principles of encoding used in MCMT are very much similar to these we consider in the present paper,  
the verification procedure is quite different. The core algorithm of MCMT relies on a  symbolic backwards reachability procedure, 
in which  first-order formulae are used for the symbolic representation of the sets of configuration. During the execution the 
reachability procedure may call the external logic engine (SMT solver) multiple times, up to several hundreds for  some examples    
as reported in \cite{MCMT}.  In the FCM method we presented here the verification procedure is much simpler and  is just 
a reduction (or compilation) to a 
\emph{single} problem in logic, which then is resolved via single call to the external logic engine (finite model builder).

In a more general context, the work  we present in this paper is related to the concepts of \emph{proof by consistency} \cite{pbc}, 
and  \emph{inductionless induction} \cite{ii} and can be seen as an investigation into the power of these concepts in 
the particular setting of the verification of parameterized systems via finite countermodel finding.

%\section{Practicalities} 

%For an set $L \subseteq \Sigma^{\ast}$ and a relation $R \subseteq \Sigma \times \Sigma$ denote by $R(L)$ the set $\{w 
%\in \Sigma \mid \exists w' \in L: (w',w) \in R\}$. Let $id \subseteq \Sigma^{\ast} \times \Sigma^{\ast}$ be the identity 
%relation and $\circ$ the composition of relations 

\section{Conclusion}~\label{sec:con}

\noindent 
We have shown how to apply generic finite model finders in the parameterized verification of linear arrays of finite automata 
models, have demonstrated the relative completeness of the method,  and have illustrated its practical efficiency. 
Further, we have shown that the verification via finite countermodel finding is at least as powerful as the 
standard regular model checking for the verification of safety properties. 
Inspection of the proofs of relative completeness  reveals that the key reason   for the 
completeness is  the existence of  regular sets separating the reachable and bad (unsafe) states. 
We conclude with the very general claim that, for any parameterized system, for which there exists a regular    
set separating reachable and unsafe states, its correctness can be demonstrated by a finite countermodel finding method.
Formal instantiations of this claim for particular classes of problems remains  a subject of ongoing and future work. 
In particular, the extension of the results presented in this paper to the case of {\em tree} regular model checking looks 
quite straightforward. More speculative and intriguing is a possibility to use {\em infinite} model building procedures \cite{Model}
for parameterized verification.     Further investigation of practical efficiency and scalability  of the method is also an important direction for future work.

\section{Acknowledgments}
The author is grateful to Michael Fisher for the helpful suggestions on this paper.

%Unlike approaches used for the similar class of problems  e.g. in \cite{Mon,Aut} our 
%method is purely \emph{reductionist} - the verification task is reduced to a disproving a formula, which then 
%delegated to available automated tools. Potential advantage of our method is its \emph{modularity} - 
%any progress in finite (and even infinite \cite{Model}) model finding procedures can be    
%incorporated immediately to the verification procedure. 

%Very recently, in \cite{Gut} the verification method using the modelling of the 
%execution of security protocols via first-order derivations and countermodel finding was proposed. This method uses the same idea as we presented here, but 
%applied to a different class of protocols, and specialized model building procedure is proposed. 

%\section*{Appendix} 

\end{document}

%% file: tempdef.tex
\newcommand{\gM}{\mathfrak{M}}

%% file: arxiv211110.bbl
\begin{thebibliography}{10}

{\small

\bibitem{Mon} Parosh Aziz Abdulla, Giorgio Delzanno, Noomene Ben Henda, Ahmed Rezine.
\newblock  Monotonic Abstraction: on Efficient Verification of Parameterized Systems. 
\newblock {\em Int. J. Found. Comput. Sci.} 20(5): 779-801 (2009)

\bibitem{context09} Parosh Aziz Abdulla, Girogio Delzanno, Ahmed Rezine. 
\newblock Approximated Context-Sensitive Analysis for Parameterized Verification. 
\newblock Lecture Notes in Computer Science, 2009, Volume 5522, 41-56


\bibitem{Ab} Parosh  Aziz Abdulla,  Jonsson B.
\newblock Verifying programs with unreliable channels. 
\newblock {\em Information and Computation}, 127(2):91-101, June 15, 1996. 



\bibitem{RMCs}
Parosh Aziz Abdulla, Bengt Jonsson, Marcus Nilsson, and Mayank Saksena, A Survey of Regular Model Checking, In Proc. of CONCUR'04, volume 3170 of LNCS, pp 35--58, 
2004. 


\bibitem{Model} 
R.~Caferra, A.~Leitsch, N.~Peltier, {\em Automated Model Building}, Applied Logic Series, 31,  Kluwer, 2004. 


\bibitem{ii}
H.~Comon. Inductionless induction. 
In R.~David, ed. \emph{2nd Int. Conf. in Logic for Computer Science: Automated Deduction. Lecture Notes}, 
Chambery, Uni de Savoie, 1994.  

\bibitem{Del03}
G.~Delzanno.
\newblock Constraint-based Verification of Parametrized Cache Coherence
  Protocols.
\newblock {\em Formal Methods in System Design}, 23(3):257--301, 2003.
\bibitem{beyond} Dana Fisman and Amir Pnueli,  Beyond Regular Model Checking,  In Proc. of FSTTCS'01, volume 2245 of LNCS, 2001.  


\bibitem{MCMT} S.Ghilardi and S.Ranise.
\newblock   MCMT: A Model Checker Modulo Theories. 
\newblock Lecture Notes in Computer Science, 2010, Volume 6173/2010, 22--29. 


\bibitem{GNRZ} S.Ghilardi, E.Nikolini, S.Ranise and D.Zucchelli. 
\newblock Towards SMT Model-Checking of Array-based Systems. 
\newblock In IJCAR, LNCS, 2008 



\bibitem{GL08} J.~Goubault-Larrecq.  Towards producing formally checkable security proofs, automatically.
In: Computer Security Foundations (CSF), pp. 224–-238 (2008)

\bibitem{GL09} J. Goubault-Larrecq. "Logic Wins!". In ASIAN'09, LNCS 5913, pages 1-16. Springer, 2009.



\bibitem{Gut} Joshua Guttman,  Security Theorem via Model Theory, arXiv:0911.2036, November 2009. 



\bibitem{HV07}   Peter Habermehl, Tomas Vojnar, Regular Model Checking Using Inference of
Regular Languages, Electronic Notes in Theoretical Computer Science (ENTCS) , 
Volume 138 , Issue 3 (December 2005) , pp 21--36, 2005 







\bibitem{JW09}
J.~Jurjens and T.~Weber, Finite Models in FOL-Based
Crypto-Protocol Verification,  P. Degano and L. Vigan`o (Eds.): ARSPA-WITS 2009, LNCS 5511, pp. 155–-172, 2009.


\bibitem{pbc} D.~Kapur and D.R.~Musser. Proof by consistency. \emph{Artificial Intelligence}, 31:125--157, 1987. 

\bibitem{Avocs09} A.~Lisitsa
\newblock Reachability as deducibility, finite countermodels and verification. 
\newblock In preProceedings of AVOCS 2009, 
Technical Report of Computer Science, Swansea University, CSR-2-2009, pp 241-243. 


\bibitem{AL09}
A.~Lisitsa
\newblock Verfication via countermodel finding\\
\verb"http://www.csc.liv.ac.uk/~alexei/countermodel/"


\bibitem{AL10} A.~Lisitsa
\newblock Reachability as deducibility, finite countermodels and verification. 
\newblock A conference version of \cite{Avocs09}, 14pp,  
\newblock (to appear in Proc. ATVA2010)


\bibitem{McCune}
W.~McCune
\newblock Prover9 and Mace4
\verb"http://www.cs.unm.edu/~mccune/mace4/"


\bibitem{RMC} M.~Nilsson. Regular Model Checking. Acta Universitatis Upsaliensis. Uppsala
Dissertations from the Faculty of Science and Technology 60. 149 pp. Uppsala. ISBN
91-554-6137-9, 2005. 

%\bibitem{esparza99verification} J.~Esparza, A.~Finkel, and R.~Mayr.
 % \newblock {On the Verification of Broadcast Protocols}.  \newblock
  %In {\em Proc.\ 14th {IEEE} Symp.\ Logic in Computer Science
  %  ({LICS})}, pages 352--359. IEEE CS Press, 1999.


\bibitem{S01} P.~Selinger,  Models for an adversary-centric protocol logic. Electr. Notes Theor.
Comput. Sci. 55(1) (2001)

%\bibitem{Aut}A.~Prasad Sistla and V.~Gyuris,  Parameterized Verification of Liner Networks using Automata as Invariants, 
%Formal Aspects of Computing, Volume 11, Number 4, 1999, pp 402-425. 

\bibitem{Weid99}C.~Weidenbach, Towards an Automatic Analysis of Security
Protocols in First-Order Logic, in H. Ganzinger (Ed.): CADE-16, LNAI 1632, pp. 314--328, 1999.
%Springer-Verlag Berlin Heidelberg 1999


}
\end{thebibliography}
